\documentclass[manuscript,trackchanges]{aastex6}
\usepackage{natbib}
\usepackage{placeins}
\usepackage{hyperref}
\usepackage{rotfloat}
\usepackage{figs}
\turnoffedit

\newcommand{\tMASS}{{\it 2MASS}}
\newcommand{\WISE}{{\it WISE}}
\newcommand{\Spitzer}{{\it Spitzer}}
\newcommand{\SOFIA}{{\it SOFIA}}

\makeatletter
\def \blfootnote{\xdef\@thefnmark{}\@footnotetext}
\makeatother

\begin{document}
\title{Massive Star Formation in the LMC. I. N159 and N160 Complexes}
\author{\vspace{-1cm}Michael S. Gordon, Terry J. Jones, Robert D. Gehrz}
\affil{Minnesota Institute for Astrophysics, School of Physics and Astronomy\\116 Church St SE, University of Minnesota, Minneapolis, MN 55455, USA}
\and
\author{\vspace{-1cm}L. Andrew Helton}
\affil{USRA--SOFIA Science Center, NASA Ames Research Center, Moffett Field, CA 94035, USA}

\begin{abstract}
  We present images and spectral energy distributions (SEDs) of massive young stellar objects (YSOs) in three star-forming \ion{H}{2} regions of the Large Magellanic Cloud: N159A, N159 Papillon, and N160. We use photometry from \SOFIA/FORCAST at 25.3--37.1~\micron\ to constrain model fits to the SEDs and determine luminosities, ages, and dust content of the embedded YSOs and their local environments. By placing these sources on mid-infrared color-magnitude and color-color diagrams, we analyze their dust properties and consider their evolutionary status. Since each object in the FORCAST images has an obvious bright near-infrared counterpart in \Spitzer\ Space Telescope images, we do not find any evidence for new, very cool, previously-undiscovered Class~0 YSOs. Additionally, based on its mid-infrared colors and model parameters, N159A is younger than N160 and the Papillon.  The nature of the first extragalactic protostars in N159, P1 and P2, is also discussed.
\end{abstract}
\keywords{\vspace{-2mm}\ion{H}{2} regions---Magellanic Clouds---stars: formation}

\section{Introduction}\label{sec:intro}

Rich with \ion{H}{2} regions and OB associations, the nearby Large Magellanic Cloud (LMC) is an excellent laboratory for the study of massive star formation. With a lower metallicity and significantly lower total galactic mass than the Milky Way, the LMC provides a unique opportunity to probe star formation under physical conditions that mirror those in dwarf galaxies and in the early universe. Stellar populations in the LMC provide tests for understanding both massive star formation and the interplay between star formation and the interstellar medium (ISM).  The conditions of star formation in the LMC have both moderate dissimilarities to the Milky Way and many similarities, such as the presence of carbon monoxide, dust, and a strong UV field.

While much work has been done on low-mass star formation in dense molecular complexes \citep{allen_2007,galametz_2009}, massive star formation is still poorly understood.  Commonly forming in tight clusters of OB stars, massive stars provide strong feedback mechanisms in shaping their local environments. A single O or B-type star, with a lifetime of just $\sim$2~Myrs, can significantly disrupt the surrounding molecular gas in the region in which it is embedded. Feedback mechanisms such as photoionization and momentum-driven winds can quench star formation for a million years or more \citep{dale_2013_3,dale_2013_4}.

Several regions of massive star formation in the LMC that have been heavily studied are the \ion{H}{2} region complexes stretching south from 30 Doradus. From north to south, these are commonly referred to as N158, N160, and N159 (see Figure~\ref{fig:doradus}). N159 hosts the first extragalactic ``protostars'' (P1, \citealt{gatley_1981}; and P2, \citealt{jones_1986}), the first extragalactic Type I OH maser \citep{caswell_1981}, and an H$_2$O maser \citep{scalise_1981}---all indicators of recent star formation. N159 lies on a peak in the CO distribution in the LMC, with another peak just south of N159, an area that has yet to form stars \citep{johansson_1998,jones_2005,mizuno_2010}. This complex has been extensively studied in the near-infrared (NIR) with the \Spitzer\ Space Telescope \citep{jones_2005,indebetouw_2008,chen_2010}, in the far-infrared (FIR) and sub-mm using the ESA {\it Herschel Space Observatory} \citep{seale_2014} and The Large Apex BOlometer CAmera \citep[LABOCA;][]{galametz_2013}, and numerous other methods.

\figDoradus

The complete absence of M supergiants indicates that, unlike in 30 Dor, star formation in N159 is very recent, with no significant epoch of star formation earlier than a few million years \citep{mcgregor_1981}. \edit1{The string of \ion{H}{2} regions stretching south from 30 Dor along the molecular ridge is suggestive of sequential star formation. Although sequential star formation has been demonstrated in some specific regions of the LMC such as N11 \citep{barba_2002}, N159 is over 100~pc south of N160, which in turn is over 150~pc south of N158, a much larger physical scale than for N11. \cite{deboer_1998} suggest that this string of \ion{H}{2} regions is due to bow-shock induced star formation as the LMC moves through the Galactic Halo. In this picture, there an age gradient from south (youngest) to north (oldest), but not due to stochastic self-propagating star formation proceeding from north to south.}

\edit1{\cite{testor_1998} reveal two OB associations in N158.  The northern association, LH~101, consists of at least two stellar populations with ages 2--6~Myrs and $\lesssim2$~Myrs, while the southern LH~104 region appears to contain only the older population. While these discoveries appear to favor a NE--SW propagation of triggered star formation, the optical spectroscopic surveys of N159 and N160 by \cite{farina_2009} show three phases of stellar evolution in a fairly uniform spread between the two \ion{H}{2} regions. It is possible that star formation proceeds at different rates among subregions; however, \cite{farina_2009} suggest a common time origin for recent star formation as a whole for N159 and N160, rather than a specific age gradient as seen by \cite{testor_1998} in N158.}

N159, presumably the youngest in the string, shows evidence for an internal triggered star formation event within the last 2 million years \citep{jones_2005}. Most Galactic \ion{H}{2} regions tend to have dense luminous cores. N159, however, is composed of at least four \ion{H}{2} regions separated from one another by 10--20~pc, each of which is roughly equivalent in luminosity to 3/4 of an M17 \ion{H}{2} region \citep{kleinmann_1973,nielbock_2001}. These are the Papillon \citep[N159-5;][]{heydari_1999}, N159A, N159AN, and the central radio peak, labeled in Figure~\ref{fig:n159}.

\figN159

The stellar population in N160 is even more evolved, with the parent molecular clouds cleared out by massive star formation \citep{oliveira_2006,galametz_2013}.  \cite{nakajima_2005} showed that several optical stellar clusters and associations forming in both N159 and N160 were likely triggered by a supergiant shell, SGS 19, just northeast of N160.  In this scenario, supported by \textit{HST} imaging from \cite{heydari_2002} and spectroscopic observations from \cite{farina_2009}, star formation in both \ion{H}{2} regions is stimulated by complex feedback interactions from 30 Doradus, SGS 19, and possibly from a nearby X-ray--emitting hot gas cloud \citep[$10^6$ K;][]{points_2000} centered at ($\alpha_{2000}$, $\delta_{2000}$)$\sim$(05:41:50, -69:42:00) \citep{nakajima_2005}.

Recently, \cite{bernard_2016} conducted high spatial-resolution $JHK_s$-band imaging of N159 with the {\it Gemini South} Adaptive Optics Imager \citep[GSAOI;][]{mcgregor_2004} on the Gemini Multi-conjugate adaptive optics System \citep[GeMS;][]{rigaut_2014,neichel_2014}. They identified several infrared-bright sources, each active sites of massive star formation.  In the NIR, these sources appear to be compact clusters, each hosting multiple components. However, they are unresolved in mid-infrared images. \Spitzer\ observations from \cite{chen_2010} of N159 and N160 as well as the GeMS images from \cite{bernard_2016} of N159A both suggest that the stars present in each cluster are at different evolutionary stages. These clusters appear to host varied stellar populations of massive young stellar objects (YSOs) and main-sequence O- and B-type stars.

While the multiple sources in each compact cluster can be resolved in the optical and near-infrared, it is not possible to resolve these individual sources with \Spitzer\ and {\it Herschel} mid- and far-infrared photometry due to the low spatial resolution at 24~\micron\ and longer wavelengths. \cite{galametz_2013} showed that the unresolved sources in N159 and N160 imaged with \Spitzer/MIPS \citep{rieke_2004} and LABOCA with a large aperture radius (100\arcsec) are radiating strongly in the mid-IR (for N160: 145~Jy at 24~\micron, 10~Jy at 100~\micron). Spectral energy distributions (SEDs) constructed from this photometry can be questionable for unresolved multiple sources in the large aperture.  However, the high-resolution $JHK_s$ images from \cite{bernard_2016} of sources in N159 reveal that the dominate component in the mid-infrared SEDs of the compact clusters is a single, massive YSO.  \cite{chen_2010} suggest that, while the optical and near-infrared emission will by dominated by exposed stars and/or older protostars that have shed their dust envelopes, model fits to the mid-infrared SED will still provide valuable information on the single embedded YSO.

In this study, we present SEDs of the massive YSOs in the three star-forming regions, N159A, N159 Papillon, and N160.  Using 25--37~\micron\ FORCAST images from the NASA Stratospheric Observatory for Infrared Astronomy \citep[\SOFIA;][]{gehrz_2009,young_2012} in combination with \Spitzer/IRAC archive photometry from 3.6--8.5~\micron\ and 100~\micron\ data from {\it Herschel}/PACS, we conduct a census of the massive YSOs and their contribution to the total luminosity of the compact clusters in the mid- to far-infrared.  Model fitting to the observed SEDs and analysis of color-magnitude diagrams enable us to characterize the luminosities, ages, and dust properties of the YSOs and their surrounding environments. Mid-infrared imaging with \SOFIA\ reveals that there are no new, very cool, previously-undiscovered YSOs present in the three \ion{H}{2} regions.

\section{Observations}\label{sec:obs}

\subsection{\SOFIA/FORCAST}\label{sec:obs_sofia}

The N159A, N159 Papillon, and N160 fields were observed with \SOFIA\ during missions OC3-D~and~F on June~19 and July~6,~2015, using the Faint Object infraRed CAmera for the \SOFIA\ Telescope \citep[FORCAST;][]{herter_2012}, which performs both imaging and spectroscopy from 5 to 40~\micron. After post-processing, FORCAST yields a pixel scale of 0\farcs768 pix$^{-1}$ providing a $3\farcm4\,\times\,3\farcm2$ effective field-of-view and 3 to 3\farcs5 spatial resolution.

We observed each of the three fields in three bands with effective wavelengths of 25.3, 31.5, and 37.1~\micron. The data from each band were taken sequentially with a mirror in place of the dichroic, as single-channel observations provide higher throughput for each channel (see \citealt{herter_2012} for a discussion of single/dual-channel mode). The bandpasses and total integration times for each filter are listed in Table~\ref{tab:obsplan}.

The chopping secondary on \SOFIA\ was configured for symmetric two-position chopping with nodding (C2N) with a chop throw of 4\arcmin\ amplitude on the sky to cancel atmospheric emission and avoid nebulosity from the \ion{H}{2} regions. The nod throw was performed parallel to the chop throw (NMC configuration), which places the array on blank sky \citep{herter_2013}. The data were calibrated and reduced with the {\it FORCAST Redux} pipeline software \cite[v1.0.6;][]{clarke_2015} at the \SOFIA\ Science Center and released to the authors as level 3 results. The WCS plate solutions for the FORCAST images were recomputed using the IRAF {\it CCMAP} program to match the astrometry to the \Spitzer/IRAC 3.6~\micron\ images. 

Image processing and aperture photometry was performed using the open-source \textit{Astropy} \citep{astropy_2013}-affiliated \textit{photutils}\footnote{\textit{photutils} provides tools for detecting and measuring photometry of astronomical sources. The software is still in development with documentation available at \url{https://photutils.readthedocs.io/}.} package. Apertures span 3 to 6\arcsec\ for sources in the three fields, chosen to encompass the extended emission of each object.  For adjacent sources with overlapping apertures, image masks were created from segmentation images generated by the \textit{photutils} implementation of the \textit{SExtractor} thresholding algorithm \citep{bertin_1996}. Background apertures were selected to represent the median nebulosity in each \ion{H}{2} region and subtracted from each target aperture. \edit1{Photometric error is measured from the uncertainty of the background in the aperture ($\sim20\%$ in all three FORCAST bands) and is included in the online version of Table~\ref{tab:phot}.}

Sources in the FORCAST images were matched with point sources in \Spitzer/IRAC images (see \S\ref{sec:obs_spitzer}). Figures~\ref{fig:duala}, \ref{fig:dualpap}, and \ref{fig:duallast} illustrate the sources analyzed in each cluster, labeled for matching to other figures and tables in this work.  Coordinates and cross-identified names to other works can be found, along with the photometry, in Table~\ref{tab:phot}.  We note that there are no sources visible in the FORCAST 25--37~\micron\ images that lack a near-infrared counterpart in the IRAC 3.6--8.0~\micron\ images. As we discuss further in \S\ref{sec:discussion}, this implies that there is no cooler, younger massive stellar population previously unaccounted for in any of the three \ion{H}{2} regions.

\begin{deluxetable}{lcccccc}
  \tablecaption{\SOFIA/FORCAST observations.\label{tab:obsplan}}
  \tablecolumns{7}
  \tablehead{\colhead{Field} & \colhead{Date} & \dcolhead{\alpha_{2000}} & \dcolhead{\delta_{2000}} & \dcolhead{\lambda_0\,(\micron)} & \dcolhead{\Delta\lambda\,(\micron)}  & \colhead{Exptime (s)}}
  \startdata
  N159A & 06/19/2015 & 05:39:40 & -69:45:45 & 25.3 & 1.9 & 420 \\
  &  &  &  & 31.5 & 5.7 & 440 \\
  &  &  &  & 37.1 & 3.3 & 470 \\
  N159 Papillon & 06/19/2015 & 05:40:05 & -69:44:45 & 25.3 & 1.9 & 420 \\
  &  &  &  & 31.5 & 5.7 & 440 \\
  &  &  &  & 37.1 & 3.3 & 470 \\
  N160 & 07/06/2015 & 05:39:45 & -69:38:30 & 25.3 & 1.9 & 520 \\
  &  &  &  & 31.5 & 5.7 & 560 \\
  &  &  &  & 37.1 & 3.3 & 460
  \enddata
  \tablecomments{Filter parameters are from the \SOFIA\ Observer's Handbook. Exposure time is calculated as the total on-source integration time from the stacked C2NC2 images. On-source integration times for various observing strategies on FORCAST are discussed here: \url{https://www.sofia.usra.edu/sites/default/files/FOR_Cyc_1_2_ExpTime.pdf}.}
\end{deluxetable}

\clearpage
\subsection{\Spitzer/IRAC}\label{sec:obs_spitzer}

Observations were made in December 2004 using all four bands of the Infrared Array Camera \citep[IRAC;][]{fazio_2004} on the \Spitzer\ Space Telescope \citep{werner_2004,gehrz_2007} as part of a Guaranteed Time Observing Program (Program ID 124) conducted by \Spitzer\ Science Working Group member R.~D.~Gehrz and his University of Minnesota team.  \cite{jones_2005} processed each post-basic-calibrated data (PBCD) image through the Spitzer Science Center Legacy MOPEX pipeline, which provides point-spread function (PSF) and aperture photometry from the Astronomical Point Source Extraction (APEX) toolkit.  For consistency with the \SOFIA/FORCAST photometry, we have updated the PSF photometry from \cite{jones_2005} for the matching FORCAST sources, using the point-response-function (PRF) methods in \textit{photutils}. The discrete PSFs for the images in the four IRAC bands were adapted from the $1/5^{\mathrm{th}}$-pixel sampled PRFs used by APEX.\footnote{Documentation on IRAC PSF/PRF available at \url{http://irsa.ipac.caltech.edu/data/SPITZER/docs/irac/calibrationfiles/psfprf/}.}

The IRAC photometry is listed in Table~\ref{tab:phot}. Note that the data presented for \tMASS\ $J$-band through IRAC 8.0~\micron\ is given in mJy, while the mid-infrared photometry is in Jy.  We estimate the photometric error in the \Spitzer/IRAC images to be $\lesssim10\%$ based on RMS variation in the background.  The positions listed in Table~\ref{tab:phot} for each object are calculated as the center-of-mass from 2D moments of the IRAC 3.6~\micron\ images.

\figDualA

\figDualPap
\clearpage

\figDualNLast

\subsection{Herschel/PACS}\label{sec:obs_herschel}
We also include in our analysis the publicly-available 100~\micron\ observations from the {\it Herschel Space Observatory} \citep{pilbratt_2010} with the PACS instrument \citep{poglitsch_2010} to constrain the Rayleigh-Jeans side of the thermal dust component of the SEDs. Observations were taken on 2010 Oct 05 as part of the HERschel Inventory of The Agents of Galaxy Evolution (HERITAGE) in the Magellanic Clouds open time key program \citep[][Obs ID 1342202224]{meixner_2013,seale_2014}. The Herschel Interactive Processing Environment (HIPE)\footnote{HIPE is a joint development by the Herschel Science Ground Segment Consortium, consisting of ESA, the NASA Herschel Science Center, and the HIFI, PACS, and SPIRE consortia.} was used to download the images, but all processing and photometry was performed using \textit{photutils} for consistency with the \SOFIA\ and \Spitzer\ data.

Since several of the unresolved sources in the 100~\micron\ images are slightly overlapping, we cannot use segmentation maps as was done for the FORCAST images to perform photometry. Instead, we scale up the PSF images of Vesta provided with HIPE (FWHM $\sim7''$). Subtracting these scaled PSFs yields masked images with most of the target profile intact. PSF photometry was then performed on the resulting source profiles with the same Vesta PSFs.  While this technique was successful for the brighter objects in the three \ion{H}{2} regions, the Papillon sources and N160 D were too faint to yield meaningful PSF photometry. For these objects, we performed aperture photometry with 8'' radius apertures.  Comparing aperture and PSF photometric values requires a correction from ``Encircled energy fraction'' (EEF) curves.\footnote{Encircled energy diagrams, color-correction strategies, and calibration factors are discussed at \url{http://herschel.esac.esa.int/twiki/pub/Public/PacsCalibrationWeb}.}  For the chosen aperture size, the EEF correction is $\sim80\%$. The width of the PACS bandpasses also requires color corrections be applied to the 100~\micron\ (green) images. Since we do not have mid-infrared spectra to convolve with the PACS response functions, we instead adopt color corrections based on the blackbody temperatures of the thermal dust component. For cool dust at 100~K, the peak of the SED will be at $\sim30$ \micron, corresponding to a color-correction calibration factor of 1.007 for the PACS green filter. We note that the color corrections for blackbody temperatures between 50 and 500~K vary only by a factor of $\sim4\%$.

\edit1{The 100~\micron\ images are shown in Figures~\ref{fig:duala}, \ref{fig:dualpap}, and \ref{fig:duallast}.  We note that every target, with the possible exception of N160 E and F, appears as a point source in the PACS images.  If each source is a compact cluster with a single YSO dominating the mid-infrared emission as \cite{chen_2010} suggest, these PACS point sources imply that the 100~\micron\ flux is emitted from each YSO envelope, rather than the average radiation field present in each \ion{H}{2} region.}

\edit1{We cross-identify our list of sources with the catalog from \cite{seale_2014}, and find matches only on N159A A and N160 A (HERITAGE sources J84.905917-69.757105 and J84.932653-69.642627, respectively).  Our 100~\micron\ photometry for N160 A is consistent with the \cite{seale_2014} measurement ($23.0\pm9$~Jy vs. $28.5\pm2$~Jy); however, we measure a flux value much higher for N159A A ($146.1\pm30$~Jy vs. $32.1\pm3$~Jy). The difference is likely due to crowding in the field, aperture size, treatment of the background, etc. Absent any other matching sources in their catalog, and to maintain consistent measurement technique in our catalog, we do not adopt their photometry for our analysis. The PACS 160~\micron\ and SPIRE $250-500$~\micron\ images lacked the spatial resolution to distinguish the target sources in the three complexes.}

\subsection{2MASS}\label{sec:obs_nearir}
\cite{jones_2005} provides $JHK_s$ photometry from the Two Micron All Sky Survey \citep[\tMASS;][]{milligan_1996} for several of the IRAC sources in N159.  We adopt those values here, and include the \tMASS\ ID in Table~\ref{tab:phot} for the matching sources.  For all other sources in N159, and every source in N160, if a match was not found in the \tMASS\ Point Source Catalog \citep{skrutskie_2006} within 0\farcs5 of the position measured in the IRAC images, we download Atlas Images from the Infrared Science Archive \citep[IRSA;][]{berriman_2008} in the three survey bands. As was done for the \SOFIA/FORCAST images, image masks were created from segmentation images, background nebulosity was measured near each source, and aperture photometry was performed with radii spanning 3 to 6\arcsec, chosen to encompass the extended emission from each object. While aperture photometry performed here is not necessarily consistent with the profile-fit photometry for sources in the \tMASS\ Point Source Catalog,\footnote{See discussion on aperture vs. profile-fit photometry and curve-of-growth corrections in the IRSA documentation at \url{http://www.ipac.caltech.edu/2mass/releases/allsky/doc/sec4_4c.html}.} the YSO age and dust properties are primarily constrained by the mid-infrared flux. As discussed in \S\ref{sec:sed}, our analysis of the SEDs is largely restricted to the mid-infrared colors (FORCAST 25.3--37.1~\micron) and PACS 100~\micron\ photometry. Any uncertainty in the near-infrared \tMASS\ photometry will have little impact on the results presented here.

While several of our sources have \WISE\ \citep{wright_2010} photometry, the angular resolution of 6 to 12\arcsec\ for the \WISE\ bands presents issues with spatially resolving, and cross-identifying, the YSOs in the three crowded fields. Thus, we do not include \WISE\ in our analysis.

\begin{sidewaystable}
  \caption{Photometry of IR Sources\label{tab:phot}}
  \scriptsize
  \begin{tabular}{cccccccccccccccc}
    \hline
    Source & 2MASS ID & Other ID\footnote{Sources labeled C2--C5 in N159A are identified as compact clusters in \cite{bernard_2016}. N159A A and N160 A are in \cite{seale_2014} (see \S\ref{sec:obs_herschel}). N160 sources are matched to sources in \cite{heydari_2002}.} & $\alpha_{2000}$ & $\delta_{2000}$ & $J$\footnote{$JHK_s$: \tMASS} & $H$ & $K_s$ & 3.6\footnote{3.6--8.0~\micron: \Spitzer/IRAC} & 4.5 & 5.8 & 8.0 & 25.3\footnote{25.3--37.1~\micron: \SOFIA/FORCAST} & 31.5 & 37.1 & 100\footnote{100~\micron: {\it Herschel}/PACS} \blfootnote{\edit1{Note: Photometric error is included in the online version of this table.}} \\
    &  &  &  &  & [mJy] & [mJy] & [mJy] & [mJy] & [mJy] & [mJy] & [mJy] & [Jy] & [Jy] & [Jy] & [Jy] \\
    \hline
    \hline
    \multicolumn{2}{c}{{\bf N159A}} \\
    A & \nodata & C5, J84.905917-69.757105 & 05:39:37.44 & -69:45:25.5 & \nodata & \nodata & \nodata & 11.75 & 23.54 & 71.88 & 217.39 & 12.09 & 27.31 & 40.89 & 149.14 \\
    B & 05393712-6945370 & \nodata & 05:39:37.03 & -69:45:36.7 & 0.96 & 2.04 & 1.35 & 7.46 & 18.48 & 41.84 & 61.26 & 2.85 & 7.38 & 9.62 & 58.93 \\
    C & 05393608-6946039 & DCL 240, C4 & 05:39:35.97 & -69:46:04.2 & 0.97 & 1.78 & 2.65 & 8.69 & 8.51 & 31.56 & 68.34 & 2.43 & 3.97 & 4.42 & 28.06 \\
    D & \nodata & C3 & 05:39:37.56 & -69:46:09.3 & \nodata & \nodata & \nodata & 17.51 & 22.65 & 81.56 & 256.64 & 10.08 & 18.32 & 24.65 & 99.53 \\
    E & 05394188-6946122 & P2, C2 & 05:39:41.80 & -69:46:11.9 & 1.22 & 2.48 & 9.12 & 43.71 & 78.13 & 154.49 & 232.54 & 3.48 & 6.75 & 9.07 & 76.5 \\
    \multicolumn{2}{c}{{\bf N159 Papillon}} \\
    A & 05400448-6944375 & MSX6C, M371 & 05:40:04.34 & -69:44:37.3 & 6.35 & 7.15 & 13.55 & 59.06 & 69.2 & 284.78 & 900.28 & 11.06 & 19.39 & 28.44 & 9.22 \\
    B & \nodata & \nodata & 05:40:09.43 & -69:44:53.9 & \nodata & \nodata & \nodata & 5.52 & 3.67 & 25.62 & 65.05 & 0.50 & 1.29 & 2.20 & 1.26 \\
    C & 05395936-6945263 & P1, M210 & 05:39:59.32 & -69:45:26.4 & 0.42 & 2.64 & 12.24 & 47.11 & 59.60 & 90.72 & 140.06 & 0.28 & \nodata & \nodata & \nodata \\
    \multicolumn{2}{c}{{\bf N160}} \\
    A & \nodata & J84.932653-69.642627, H46-48 & 05:39:43.78 & -69:38:33.4 & \nodata & \nodata & \nodata & 44.98 & 125.64 & 230.77 & 283.14 & 10.34 & 20.30 & 29.71 & 23.02 \\
    B & \nodata & H56, H58 & 05:39:46.00 & -69:38:38.9 & 20.83 & 20.46 & 31.90 & 59.68 & 87.31 & 230.05 & 600.17 & 38.99 & 69.98 & 97.46 & 40.64 \\
    C & \nodata & H36, H41 & 05:39:44.40 & -69:38:47.4 & \nodata & \nodata & \nodata & 9.14 & 31.70 & 140.40 & 477.1 & 7.45 & 12.32 & 17.27 & 20.41 \\
    D & \nodata & H35 & 05:39:43.19 & -69:38:53.9 & 17.71 & 13.38 & 20.22 & 34.58 & 68.84 & 189.28 & 593.95 & 10.76 & 14.73 & 18.75 & 6.97 \\
    E & \nodata & H24 & 05:39:38.99 & -69:39:11.7 & 2.45 & 5.51 & 7.21 & 24.65 & 27.86 & 140.24 & 380.45 & 9.91 & 21.59 & 34.51 & 19.73 \\
    F & \nodata & \nodata & 05:39:38.70 & -69:39:04.0 & \nodata & 0.49 & 2.22 & 17.20 & 15.99 & 116.55 & 339.00 & 4.65 & 8.21 & 12.46 & \nodata \\
    \hline
  \end{tabular}
\end{sidewaystable}

\clearpage

\section{Spectral Energy Distributions and YSO Modeling}\label{sec:sed}

To derive the characteristics of the resolved YSOs in the FORCAST images, we use the 2D radiative transfer YSO model grid from \cite{robitaille_2007} and their SED fitter tool to determine physical parameters of YSOs from available photometry data from \tMASS\ $J$-band through {\it Herschel}/PACS 100~\micron. The models assume an accretion scenario for the star-formation process, where the central source is surrounded by a dusty accretion disk, an infalling flattened envelope, and the presence of bipolar cavities.  We have used the command-line version of the SED fitting tool, which finds the best fitting SED (minimizing $\Delta\chi^2$ per photometric point) from $200\,000$ precomputed models.\footnote{\textit{Python} port of the original \textit{Fortran} SED fitter code \citep{robitaille_2007} found at \url{http://sedfitter.readthedocs.org/}.}

The model inputs include distance to the source, approximate foreground visual extinction (A$_{\mathrm{V}}$), the flux in each passband, and an estimate of photometric error for each filter.  For the FORCAST and PACS filters, we convolve all of the precomputed models with normalized filter response functions to extend the SED fitting tool out to 100~\micron.  While A$_{\mathrm{V}}$ can be left as a completely free parameter, we set a lower-limit of A$_{\mathrm{V}} = 4$~mag, based on the \cite{jones_2005} discussion of H$\alpha$ obscuration in N159AN. We assume no significant variation in foreground extinction across the \ion{H}{2} regions, and we note that this assumption will have no impact on the SED fitting in the mid- to far-infrared where the extinction is very low.

The output parameters from the model fitting are summarized in Table~\ref{tab:model}, and the best-fitting SEDs for selected sources in all three clusters plotted are plotted in Figure~\ref{fig:models}. Although the model SEDs are consistent with the observed flux densities, we note that the derived parameters must be treated with caution.  PAH and small-grain thermal emission, which are strong sources of mid-infrared flux in YSOs with hot stellar sources \citep{robitaille_2006,seale_2009,nandakumar_2016}, are absent in the synthetic SEDs. Considering that the FORCAST 25--37~\micron\ fluxes are the primary constraint for the models, the insufficient treatment of PAH emission creates some uncertainty in the shape of the SEDs in the mid-infrared.

Additionally, the precomputed YSO models assume that massive YSOs are simply scaled-up versions of their lower mass counterparts.  The radii, temperatures, and ages are derived from pre-main sequence isochrones formulated from models of low- to intermediate-mass stars by \cite{siess_2000}, not from isochrones that apply to the potentially massive YSOs embedded in the LMC \ion{H}{2} regions. The \cite{siess_2000} isochrones and evolutionary tracks can differ significantly from the Geneva group evolutionary models of massive stars \citep{charbonnel_1999,ekstrom_2012}, particularly for isochrones less than $10^6$ years---a timescale comparable to the lifetimes of many high-mass stars.  For example, though the SED modeling predicts an age of only $\sim30\,000$ years for cluster N159A, we would have expected ages between 0.1--0.5~Myrs based on their far-infrared colors and location on the color-magnitude diagram (see \S\ref{sec:color}; Figures~\ref{fig:HRD} and \ref{fig:twocolor}). As \cite{siess_2001} describes in a review of his earlier work on pre-main sequence evolutionary tracks (\citealt{siess_2000}; the isochrones used by the \citealt{robitaille_2007} models), much of the uncertainty in YSO ages stems from ambiguity in assigning ``time zero'' to the models, as well as treatment of rotation and convection in the stellar evolution code.  Even modest uncertainty in metallicity, luminosity, and temperature can have a significant impact on the derived ages, with age estimates varying by a factor of 2 to 4 for stars younger than 1--2~Myrs \citep{siess_2001}. Therefore, we present the model ages in Table~\ref{tab:model} with some caution.

\edit1{\cite{meynardier_2004} overlay isochrones on color-magnitude and color-color diagrams of sources in N159 Papillon using $JHK_s$ photometry taken on VLT.  While they conclude 3~Myrs is the best-fitting age for the massive star population, they note that any isochrone between 1 and 10~Myrs would be consistent with their data due to degeneracy in the near-IR colors of massive stars---a different problem than we face matching our longer wavelength SEDs with near-IR data since the VLT images in \cite{meynardier_2004} fully resolve the stellar populations in the Papillon. Their age of 3~Myrs is older than what we estimate from the YSO models; however, if internal triggered star formation is occurring within the N159 \ion{H}{2} region, then an age difference may exist between the central cluster stars and embedded YSOs at the periphery \citep[``bright-rimmed clouds,''][]{ikeda_2008,panwar_2014}. \cite{bernard_2016} additionally suggest that local density fluctations in N159 may lead to fragmented clumps collapsing at different dynamical timescales. While we cannot definitively argue for a particular scenario---or measure the ages of individual YSOs---we explore their Class 0/I/II designations in \S\ref{sec:twocolor} using mid-infrared color-color diagrams.}

To characterize the thermal dust emission, we fit a gray body spectrum to the mid- to far-infrared photometry following the SED modeling by \cite{rathborne_2010} of protostellar cores.  The emissivity is characterized by a power-law dependence $\epsilon_{\lambda}\propto\lambda^{-\beta}$. While \cite{jones_2016} find a good fit for $\beta=1.6$ for two Galactic YSOs embedded in an infrared dark cloud, we find values of $\beta$ of 1.8 for sources in N159 and and 2.0 for those in N160. These emissivity parameters are consistent with the range of $\beta$ between 1.6 and 2.1 for the embedded cores in \cite{rathborne_2010}. In Figure~\ref{fig:models} we show the gray body fits as gray dashed lines, where the fitting was restricted to $\lambda > 20~\micron$ to fit only the long-wavelength thermal dust component, along with the full YSO model SEDs in blue. The average dust temperatures for sources in N159A, N159 Papillon, and N160 are $\sim60$, 100, and 90~K, respectively. The best-fitting dust temperatures for individual sources are included in Table~\ref{tab:model}.

\edit1{For some of the sources we note a departure in the shape of the SED around the \SOFIA/FORCAST photometry from the gray body fits.  A similar trend was shown in the SEDs of infrared dark clouds MM1 and MM2 from \cite{jones_2016}, which they suggest reveals a warmer gas and dust component around Class~0~YSOs. \cite{ma_2013} perform gray body fits to Galactic protostar SEDs in the CHaMP survey using up to four separate thermal components between 30 and 240~K to study such a scenario. This functionality is not fully realized in the \cite{robitaille_2007} model grid, nor in the 3D models from \cite{whitney_2013}.  Similar SEDs are shown in \cite{chen_2010} and \cite{seale_2014}, some of which have the same bump around 20--40~\micron\ seen in MM1, MM2, and these sources. From the models shown in Figure~\ref{fig:models}, it is clear that a single thermal dust component is insufficient to fully fit the mid-infrared SED. Therefore, the best-fitting dust temperatures  in Table~\ref{tab:model} represent the \textit{coolest} dust component observed, while dust species spanning multiple temperatures may still be present in the beam. Considering that the uncertainty in the photometry is between 10 and 25\% for the 25.3--100~\micron\ data, we expect the error in the derived dust temperature to scale similarly.}

\figModels
\clearpage
\begin{deluxetable}{ccccccccc}
  \tablecaption{Derived YSO Model Parameters\label{tab:model}}
  \tablecolumns{9}
  \tabletypesize{\scriptsize}
  \tablehead{\colhead{Source} & \colhead{Mean Fgd A$_{\mathrm{V}}$\tablenotemark{a}} & \colhead{Mean Age} & \colhead{$\tau_{\,\mathrm{Ks}}$\tablenotemark{b}} & \colhead{L$_{\star}$} & \colhead{T$_{\star}$} & \colhead{T$_{\mathrm{dust}}$\tablenotemark{c}} & \colhead{M$_{\mathrm{disk}}$} & \colhead{$\dot{\mathrm{M}}_{\mathrm{disk}}$} \\
    \colhead{} & \colhead{[mag]} & \colhead{[yr]} & \colhead{}  & \colhead{[L$_{\odot}$]} & \colhead{[K]} & \colhead{[K]} & \colhead{[M$_{\odot}$]} & \colhead{[M$_{\odot}$/yr]}}
  \startdata
      {\bf N159A} & 7.4 &  $3 \times 10^4$ \\
      A & & & 31.6 & $3.9 \times 10^{5}$ & $4.3 \times 10^{4}$ & \phn60 & $3.2 \times 10^{-3}$ & $7.8 \times 10^{-6}$ \\
      B & & & 19.4 & $8.5 \times 10^{4}$ & $3.4 \times 10^{4}$ & \phn57 & $9.0 \times 10^{-3}$ & $6.0 \times 10^{-7}$ \\
      C & & & 17.0 & $1.1 \times 10^{5}$ & $3.9 \times 10^{4}$ & \phn61 & $9.9 \times 10^{-3}$ & $1.2 \times 10^{-7}$ \\
      D & & & 21.5 & $2.6 \times 10^{5}$ & $4.3 \times 10^{4}$ & \phn63 & \nodata & \nodata \\
      E & & & 13.6 & $6.0 \times 10^{4}$ & $2.8 \times 10^{4}$ & \phn66 & $8.4 \times 10^{-3}$ & $3.6 \times 10^{-7}$ \\
      {\bf N159 Papillon} & 6.1 & $7 \times 10^5$ \\
      A & & & 21.1 & $2.2 \times 10^{5}$ & $4.3 \times 10^{4}$ & 123 & $1.4 \times 10^{-2}$ & $6.6 \times 10^{-7}$ \\
      B & & & 31.9 & $1.4 \times 10^{4}$ & $3.0 \times 10^{4}$ & 105 & $1.6 \times 10^{-1}$ & $1.6 \times 10^{-7}$ \\
      C & & & \phn8.1 & $1.1 \times 10^{4}$ & $2.8 \times 10^{4}$ & \phn92 & $3.9 \times 10^{-2}$ & $5.6 \times 10^{-6}$ \\
      {\bf N160} & 4.4 & $2 \times 10^5$ \\
      A & & & 10.1 & $2.5 \times 10^{5}$ & $4.4 \times 10^{4}$ & \phn84 & $1.5 \times 10^{-1}$ & $3.5 \times 10^{-6}$ \\
      B & & & 37.2 & $3.8 \times 10^{5}$ & $4.6 \times 10^{4}$ & \phn96 & $5.8 \times 10^{-1}$ & $1.4 \times 10^{-6}$ \\
      C & & & \phn5.6 & $2.4 \times 10^{5}$ & $4.3 \times 10^{4}$ & \phn80 & \nodata & \nodata \\
      D & & & 74.6 & $1.4 \times 10^{5}$ & $4.1 \times 10^{4}$ & \nodata & $8.2 \times 10^{-2}$ & $2.8 \times 10^{-7}$ \\
      E & & & 13.6 & $2.5 \times 10^{5}$ & $4.4 \times 10^{4}$ & \phn90 & $1.3 \times 10^{-4}$ & $1.2 \times 10^{-7}$ \\
      F & & & 84.8 & $9.9 \times 10^{4}$ & $3.9 \times 10^{4}$ & \nodata & $6.4 \times 10^{-2}$ & $3.1 \times 10^{-6}$
      \enddata
      \tablenotetext{a}{Extinction in the foreground of the models averaged by cluster.}
      \tablenotetext{b}{Optical depth at 2.2~\micron\ inside the model down to the stellar surface, assuming \cite{indebetouw_2005} extinction.}
      \tablenotetext{c}{Dust temperatures calculated from fitting modified gray bodies ($\beta = 1.8$ for N159, 2.0 for N160) to fluxes for $\lambda > 20$~\micron\ (see Figure~\ref{fig:models} and \S\ref{sec:sed}).}
\end{deluxetable}

\clearpage
\FloatBarrier

\section{Discussion}\label{sec:discussion}
\subsection{Color-Magnitude Diagram}\label{sec:color}

For the SEDs in Figure~\ref{fig:models}, we note a rise in energy from the FORCAST photometry (20--40~\micron) out to 100~\micron\ in the N159A sources. In the other two clusters, the SEDs appear to peak at shorter wavelengths.  To explore this color difference among the clusters, we plot the far-infrared colors of the YSOs in each cluster against their bolometric luminosities on the color-magnitude diagram in Figure~\ref{fig:HRD}. Based on the fitting parameters in Table~\ref{tab:model}, the color difference between 31.5 and 100~\micron\ suggests that the younger YSOs in N159A are young enough to have retained much of their cooler gas and dust envelopes.  Therefore, the cooler thermal dust component dominates the mid- to far-infrared emission, which makes the sources in N159A appear redder. The older YSOs in N160 have shed their outer envelopes, revealing the warmer stellar cores. The temperature scale shown in Figure~\ref{fig:HRD} is a rough correlation between the $[31.5]-[100]$ color and the dust temperatures found from the gray-body fitting discussed in \S\ref{sec:sed}. \edit1{The temperatures describe the coolest dust in the circumstellar ejecta, while the \SOFIA/FORCAST fluxes reveal that warmer components may also be present.} Discussed further in \S\ref{sec:twocolor}, the warm temperatures for the two Papillon sources is likely a result of internal triggered star formation in N159.

\figHRD

The luminosities in Figure~\ref{fig:HRD} are calculated by integrating the SED model fits from the near- to far-infrared. This treatment of the broadband photometry implies that the \ion{H}{2} regions are composed of singular, infrared-bright sources.  However, the high-resolution $JHK_s$ images from \cite{bernard_2016} and \Spitzer\ observations from \cite{chen_2010} indicate that each infrared source is actually a compact cluster, hosting as many as 12 stars in the case of C4 (here, identified as N159A C in Figure~\ref{fig:duala}).  Since \cite{bernard_2016} showed that the sources follow a distribution in ages, the mid- to far-infrared emission will be dominated by one, or at most two, embedded YSOs.  The optical to near-infrared photometry may be contaminated by unobscured stars in addition to YSOs that have shed their dust envelopes.  Thus, the model fits to the FORCAST and PACS photometry in the mid- to far-infrared will more accurately reflect the properties of a single massive YSO, with the caveat that the near-infrared fluxes may contain contributions from several sources. Regardless, the bolometric luminosities in Figure~\ref{fig:HRD}, as well as the $[31.5]-[100]$ color, should be unaffected by multiplicity, as the SEDs in Figure~\ref{fig:models} indicate that little energy is radiated in the near-infrared relative to the thermal dust emission at wavelengths greater than 20~\micron.

Two sources in the Papillon have the same color temperature as those in N160, as well as similar dust temperatures of $\sim100$~K according to the thermal dust fitting in Figure~\ref{fig:models}.  The third source in the Papillon (labeled C in Figure~\ref{fig:dualpap} and the tables) is P1 in \cite{gatley_1981} and was postulated by \cite{jones_2005} as a Class~I~YSO.  However, with no mid-infrared emission detectable in the FORCAST 31.5 and 37.1~\micron\ images, P1 is most likely {\it not} a YSO. As suggested in \cite{jones_2005}, it may instead be a carbon star.  If true, P2 (here, N159A source E), discovered by \cite{jones_1986}, would then be the first extragalactic protostar discovered.  P2 is in \cite{bernard_2016} compact cluster C2, which GeMS images revealed to contain two red sources. \cite{chen_2010} identify a Class~I/II~YSO at this location (object 053941.89--694612.0), which they suggest may have multiple components; however, they note that only one source is likely dominating the emission in the mid-infrared \citep[source 123 from][with VLT/NACO observations]{testor_2006}.

Due to deep obscuration in the optical and near-infrared, \cite{bernard_2016} could not resolve any stars in their compact cluster C5 (here, N159A source A). However, \cite{chen_2010} identify an embedded YSO, object 053937.56--694525.4, which they determine is a Class~I with an inferred spectral type of O6 V based on its luminosity from SED fitting. More recently, \cite{fukui_2015} detected a complex filamentary structure of dense CO gas using ALMA at this location, and conclude that their YSO-N is Class~0/I with a mass of $\sim31~M_{\odot}$.

\clearpage

\subsection{Color-Color Diagram}\label{sec:twocolor}
As discussed in \S\ref{sec:sed}, the ages derived from the SED fitting can be somewhat questionable for these high-mass protostars.  Rather than definitively categorizing these sources based on model ages, we instead place them on a color-color diagram to qualitatively study the YSOs by \ion{H}{2} region. In Figure~\ref{fig:twocolor}, we plot sources by IRAC and FORCAST colors, with the YSO classification scheme from \cite{reach_2004} and \cite{rho_2006}. While \cite{reach_2004} originally generated the Class 0/I and II regions based on \Spitzer/MIPS 24~\micron\ photometry, we substitute the FORCAST 25.3~\micron\ data since MIPS lacked the spatial resolution for our cluster sources. The MIPS 24~\micron\ filter is wider than the FORCAST filter suite,\footnote{MIPS spectral response falls to 50\% beyond 26~\micron.} so we note that this color substitution is justified for comparison to previous work. In this color-color space, main-sequence stars or unobscured photospheres would fall near the lower-left ($[8.0]-[25.3] = [3.6]-[5.8] = 0$).  Class~0 to Class~I protostars are the reddest sources, with a rise in their SEDs through the mid-infrared due to cool thermal dust emission.  Class~II YSOs, generally young objects with disks, will show an infrared excess relative to stellar photospheres in the near-infrared as they have not yet completely shed their outer envelopes.  Since the foreground extinction in the near-infrared is not well measured, the classification of sources near the boundary between the two marked regions in Figure~\ref{fig:twocolor} is uncertain.

\cite{rho_2006} suggest three mechanisms for producing the ``hot excess'' region of their color-color diagram. These sources could be pre-main sequence Herbig Ae/Be stars with hot circumstellar dust creating an IR excess (as opposed to post-main sequence classical Be stars whose near-infrared excess is due to free-free emission). Indeed, \cite{indebetouw_2004} and \cite{nakajima_2005} identified several objects in the N159/N160 as Herbig Ae/Be stars, including one object in N159A~C (C4 in \citealt{bernard_2016}, 053935.99--694604.1 in \citealt{chen_2010}), based on the \Spitzer/IRAC colors.

\figTwoColor

Alternatively, \cite{rho_2006} propose ``hot excess'' protostars as Class~0/I YSOs with more active accretion than is typical, which would produce a brighter, hot component ($\sim500$ K) in the near-infrared SED. However, the derived YSO parameters in Table~\ref{tab:model} from SED fitting suggest that while the Papillon sources have a higher disk mass than those in N159A, the masses and accretion rates are consistent with sources in N160, which do not appear as ``hot excess'' sources in the color-color diagram.

Finally, \cite{rho_2006} suggest a possible connection between the ``hot excess'' and star-forming environments in high UV radiation fields.  While \cite{rho_2006} were studying Galactic YSOs in the Trifid Nebula (M20), \cite{heydari_1999} discovered that YSOs in the Papillon \citep[and in N160;][]{heydari_2002} were embedded in what they call high-excitation blobs (HEBs).  HEBs are characterized by large [\ion{O}{3}]~$\lambda~5007$/H$\beta$ line ratios, which suggest a hard radiation field capable of doubly-ionizing almost all the oxygen atoms in the Papillon \citep{heydari_1999}.  With relatively high foreground extinction in the direction of the Papillon ($\mathrm{A}_\mathrm{V}\gtrsim6$), the gas and dust surrounding the embedded protostars is at a high enough density to prevent UV leakage, which implies that the \ion{H}{2} region is thermalizing its UV radiation field. This energy would appear as a hot excess in the near-infrared SEDs of the YSOs, driving these sources to the left in Figure~\ref{fig:twocolor}.

\section{Conclusions}\label{sec:conclusions}

In this study we have provided a census of extremely young, luminous YSOs in three LMC \ion{H}{2} regions. From 25--37~\micron\ imaging with \SOFIA/FORCAST and near-infrared observations with \Spitzer/IRAC and \tMASS, to the far-infrared with {\it Herschel}/PACS, we construct SEDs for each of the massive YSO candidates in N159A, N160, and the Papillon. While the isochrones used by the ~\cite{robitaille_2007} YSO models provide uncertain ages for massive stars, the positions of the YSOs on the far-infrared color-magnitude diagram indicate increased dust obscuration in N159A sources, which implies that N159A is indeed younger than N160. \edit1{From these data, though, we cannot conclude whether there is a definitive age gradient progressing N--S, or if N159 and N160 share a common time origin as \cite{farina_2009} suggest. However, we note that \cite{deboer_1998} lists ages that support the age gradient scenario between 30 Dor and N159.}

The Papillon, containing sources with a hotter $[8.0]-[25.3]$ color and infrared excess in the IRAC bands, is consistent with the \cite{heydari_1999} description of a high-excitation blob in a hard radiation field. This scenario is supported in the models: the dust temperatures and larger disk mass estimated from the model fitting and the warmer dust temperatures from the gray-body fitting suggest that the Papillon YSOs have shed some portion of their outer dust envelopes. The exposed, hotter disk components dominate the mid-infrared SEDs of the Papillon sources. \edit1{If N159 has experienced a round of internal triggered star formation, as \cite{jones_2005} suggests, the resulting hard radiation field may explain why the Papillon sources appear older than the YSOs in N159A. In contrast, \cite{chen_2010} concludes that N159-W (which encompasses N159A) more likely formed spontaneously, while N159-E (which encompasses the Papillon) was triggered by the \ion{H}{2} region expanding into the molecular cloud to the east.}

We note that the near-infrared photometry is possibly contaminated by other stars in the compact clusters. While the far-infrared color used in the color-magnitude diagram in Figure~\ref{fig:HRD} is diagnostically useful for characterizing a single YSO in each cluster, the IRAC colors in Figure~\ref{fig:twocolor} should be viewed with caution. The Galactic YSOs observed by \cite{reach_2004} and \cite{rho_2006} were spatially resolved as single objects, and so applying their near-infrared color criteria and classification scheme to sources with possible multiplicity is uncertain.

The spatial resolution of FORCAST in the mid-infrared nonetheless provides a complete catalog of embedded protostars in these LMC \ion{H}{2} regions. Since each object in the FORCAST images has an obvious bright near-infrared counterpart in \Spitzer, we do not find any evidence for new, very cool, undiscovered Class~0 YSOs. The \SOFIA/FORCAST observations support the conclusions from \cite{chen_2010} and \cite{bernard_2016} that only one massive YSO is present in each compact cluster, and that this single object dominates the mid- to far-infared component of the SED.

Finally, we conclude that since P1 is lacking a cold dust envelope, it is likely not a YSO, and perhaps a carbon star as suggested in \cite{jones_2005}. This observation would make P2 the first extragalactic protostar to have been discovered by \cite{jones_1986}.

We note that N158, the northernmost and oldest region in the string of LMC star-forming environments, is potentially evolved enough to have triggered a significant second generation of massive stars.  Mid-infrared observations of this \ion{H}{2} region will be discussed in Paper~II of this series.\\

\acknowledgements
Research by M.~S.~Gordon, T.~J.~Jones, and R.~D.~Gehrz on star formation in the local universe is supported by NASA through award \# SOF 03-0049 and the United States Air Force. The authors thank R.~M.~Humphreys and M.~W.~Werner for helpful discussion.

This publication makes use of data products from the Two Micron All Sky Survey, which is a joint project of the University of Massachusetts and the Infrared Processing and Analysis Center/California Institute of Technology, funded by the National Aeronautics and Space Administration and the National Science Foundation, as well as observations made with the NASA/DLR Stratospheric Observatory for Infrared Astronomy (SOFIA). \SOFIA\ is jointly operated by the Universities Space Research Association, Inc. (USRA), under NASA contract NAS2-97001, and the Deutsches SOFIA Institut (DSI) under DLR contract 50 OK 0901 to the University of Stuttgart. This work is based in part on archival data obtained with the Spitzer Space Telescope, which is operated by the Jet Propulsion Laboratory, California Institute of Technology under a contract with NASA. The Herschel spacecraft was designed, built, tested, and launched under a contract to ESA managed by the Herschel/Planck Project team by an industrial consortium under the overall responsibility of the prime contractor Thales Alenia Space (Cannes), and including Astrium (Friedrichshafen) responsible for the payload module and for system testing at spacecraft level, Thales Alenia Space (Turin) responsible for the service module, and Astrium (Toulouse) responsible for the telescope. PACS has been developed by a consortium of institutes led by MPE (Germany) and including UVIE (Austria); KU Leuven, CSL, IMEC (Belgium); CEA, LAM (France); MPIA (Germany); INAF-IFSI/OAA/OAP/OAT, LENS, SISSA (Italy); IAC (Spain). This development has been supported by the funding agencies BMVIT (Austria), ESA-PRODEX (Belgium), CEA/CNES (France), DLR (Germany), ASI/INAF (Italy), and CICYT/MCYT (Spain).

{\it Facilities:} {\it Herschel}/PACS, \SOFIA/FORCAST, \Spitzer/IRAC

\clearpage
\bibliography{cite}

\end{document}